\documentclass[%
 reprint,
 amsmath,amssymb,
 aps,
]{revtex4-1}
\usepackage{graphicx}% Include figure files
\usepackage{dcolumn}% Align table columns on decimal point
\usepackage{bm}% bold math
\begin{document}

\preprint{APS/123-QED}

\title{\boldmath Measuring the Fourth Generation $b \to s$ Quadrangle at the LHC}

\author{Wei-Shu Hou$^{a,b}$, Masaya Kohda$^{a}$, and Fanrong Xu$^{a}$}
 \affiliation{$^{a}$Department of Physics, National Taiwan University,Taipei, Taiwan 10617\\
$^{b}$National Center for Theoretical Sciences, National Taiwan University, Taipei, Taiwan 10617}%Lines break automatically or can be forced with \\

%\date{\today}% It is always \today, today,
             %  but any date may be explicitly specified

\begin{abstract}
We show that simultaneous precision measurements of the
$CP$-violating phase in time-dependent $B_s \to J/\psi\phi$ study
and the $B_s \to \mu^+\mu^-$ rate,
together with measuring $m_{t'}$ by direct search at the LHC,
would determine $V_{t's}^*V_{t'b}$ and therefore
the $b\to s$ quadrangle in the four-generation standard model.
The forward--backward asymmetry in $B\to K^*\ell^+\ell^-$
provides further discrimination.
\begin{description}
%\item[Usage]
% Secondary publications and information retrieval purposes.
\item[PACS numbers]
14.65.Jk % Other quarks (e.g., 4th generations)
12.15.Hh % Determination of Cabibbo-Kobayashi & Maskawa (CKM) matrix elements
11.30.Er % Charge conjugation, parity, time reversal, and other discrete symmetries
13.20.He % Decays of bottom mesons (Leptonic,semileptonic, and radiative decays of mesons)
%\item[Structure]
% You may use the \texttt{description} environment to structure your abstract; use the optional argument of the \verb+\item+ command to give the category of each item.
\end{description}
\end{abstract}

\pacs{Valid PACS appear here}% PACS, the Physics and Astronomy
                             % Classification Scheme.
%\keywords{Suggested keywords}%Use showkeys class option if keyword
                              %display desired
\maketitle

%\tableofcontents

\section{\label{sec:Intro}INTRODUCTION\protect\\}

Much like the completion of the three-generation
``$b\to d$ triangle" in 2001 by the B factories,
we may be at the dawn of measuring the ``$b\to s$ quadrangle"
at the LHC, {\it if} a fourth generation of quarks should exist.

Measurement of the time-dependent $CP$-violating (CPV) phase
$\sin2\beta/\phi_1$ in $B_d \to J/\psi K^0$ decays by the
BaBar and Belle experiments confirmed~\cite{PDG} the
Kobayashi--Maskawa~\cite{Kobayashi:1973fv} mechanism
of the standard model with three generations of quarks (SM3).
Here, $\sin2\beta = \sin2\phi_1 \equiv \sin2\Phi_{B_d}$ is the
CPV phase of the $\bar B_d \to B_d$ mixing amplitude.
With the continuous run of the Large Hadron Collider (LHC)
throughout 2011-2012, the LHCb experiment will measure
$\sin2\Phi_{B_s}$, the CPV phase of $\bar B_s \to B_s$ mixing,
via time-dependent study of $B_s\to J/\psi\phi$ and similar decays.
We point out that, together with the measurement of
$B_s\to \mu^+\mu^-$ rate, which is accessible not only by
LHCb, but by the CMS experiment (and eventually, ATLAS) as well,
combined with the direct search program of fourth-generation quarks,
one may determine the Cabibbo--Kobayashi--Maskawa (CKM) mixing matrix
element~\cite{Kobayashi:1973fv, Cabibbo:1963yz, Glashow:1970gm}
product $V_{t's}V_{t'b}^*$, thereby complete the SM4 quadrangle of
\begin{equation}
 V_{us}V_{ub}^* + V_{cs}V_{cb}^* + V_{ts}V_{tb}^* + V_{t's}V_{t'b}^* = 0.
\label{bsQuad}
\end{equation}
Much progress has been made in summer 2011 on the above,
so let us retrace how we reached the present.

Interest in the fourth generation renewed with the
``$B\to K\pi$ direct CPV (DCPV) difference" puzzle:
DCPV in $B^+ \to K^+\pi^0$ and $B^0 \to K^+\pi^-$
appeared opposite in sign~\cite{BelleNature, PDG},
even though they proceed by the same spectator diagrams.
The effect could be due to~\cite{HNS}
the nondecoupling of the heavy SM4 $t'$ quark in
the $bsZ$ penguin, which brings in
a new CPV phase in $V_{t's}^*V_{t'b}$.
But hadronic effects make the $B\to K\pi$
DCPV measurements less amenable to interpretation.

However, an SM4 effect in the $b\to s$ $Z$-penguin loop should give
a correlated effect in the $b\bar s \to s\bar b$ box diagram,
making $\sin2\Phi_{B_s}$ large and negative~\cite{HNS, Hou:2005yb},
in contrast with $-0.04$ in SM3.
After the 2006 measurement~\cite{PDG} of $B_s$ mixing,
i.e., $\Delta m_{B_s}$, by the CDF experiment at the Tevatron,
the ``prediction" was strengthened~\cite{Hou:2006mx} to
``$\sin2\Phi_{B_s} = -0.5$ to $-0.7$ for $m_{t'} = 300$ GeV."
Interestingly, by 2008, both the CDF and D0 experiments reported~\cite{PDG}
hints for negative $\sin2\Phi_{B_s}$
(called respectively $-\sin2\beta_s$ and $\sin\phi_s$).
Although weakening in 2010, the measurement~\cite{LHCb10}
by LHCb using just the 2010 data of 36 pb$^{-1}$ showed
a $\sin\phi_s$ that deviated from SM3 by $1.2\sigma$,
\emph{i.e., in same direction as CDF and D0!}
So, there was much anticipation for LHCb to unveil their
result with 10 times the data.
To one's surprise, however, analyzing 0.34 fb$^{-1}$ data,
the LHCb experiment found~\cite{Raven}
\begin{equation}
 \phi_s \equiv 2\Phi_{B_s} = 0.03 \pm 0.16 \pm 0.07,\ \ \
  {\rm (LHCb\ 0.34\ fb}^{-1})
\label{phis1108}
\end{equation}
which is consistent with zero (hence SM3).
In fact, $B_s\to J/\psi\phi$ alone gave $0.13 \pm 0.18 \pm 0.07$,
while Eq.~(\ref{phis1108}) is the combined result with
$B_s\to J/\psi f_0(980)$.

There was another development that aroused the interest
in the fourth generation in the past few years, namely
the realization~\cite{Kribs:2007nz,Holdom:2009rf} in 2007
that electroweak precision tests did not firmly rule out
a fourth generation, but rather indicated that
the $t'$, $b'$ quarks be heavy, split in mass --- but
not by too much --- while the Higgs mass bound would loosen.
The direct search for $t'$ and $b'$ at the Tevatron had in any case
been ongoing. At the LHC, the limit~\cite{Chatrchyan:2011em} of
$m_{b'} > 361$ GeV (95\% C.L.) was reached with 2010 data alone,
and became 495 (450) GeV for $b'$ ($t'$) by~\cite{DeRoeck} summer 2011.
We are already at the doorstep of the unitarity bound (UB) of
500--550 GeV~\cite{Chanowitz:1978uj}.

It is difficult to enhance $B_s \to \mu^+\mu^-$ in SM4
by more than a factor of 2,
because it is constrained by $B \to X_s\ell^+\ell^-$,
which is consistent with SM3 in rate.
Hence, this mode appeared less relevant for SM4, until recently.
Based on 2010 data, the competitive limit~\cite{Aaij:2011rj}
by LHCb was already within 20 times the SM3 expectation
of $3.2 \times 10^{-9}$~\cite{Buras:2010pi}.
%, and with 1 fb$^{-1}$ data, a 3$\sigma$ measurement can
%be made~\cite{Teubert} if ${\cal B}(B_s \to \mu^+\mu^-) = 7\times 10^{-9}$.
%
Since 2010, the progress is significant, both at the Tevatron and the LHC
 (see \textbf{Note Added}),
and a measurement of $B_s \to \mu^+\mu^-$ at the SM3 level
now seems possible with 2011-2012 LHC data.
With the signal of two charged tracks from a displaced vertex,
the CMS experiment has demonstrated its competitiveness,
in part due to an advantage in luminosity.
%
%The drama of summer 2011 is that, just after CDF reported~\cite{CDFmumu11}
%a hint, ${\cal B}(B_s\to \mu^+\mu^-) = (18^{+11}_{\;-9})\times 10^{-9}$,
%it was challenged by both LHCb~\cite{LHCb-mumu11} and CMS~\cite{CMSmumu11} within a week.
The combined result~\cite{comboBsmumu} of LHCb and CMS gives
\begin{equation}
 {\cal B}(B_s\to \mu^+\mu^-) < 11 \times 10^{-9},\ \
  {\rm (LHCb+CMS,\,2011)}
\label{Bsmumu1107}
\end{equation}
at 95\%\ CL, which is only 3.5 times the SM3 level.

\begin{figure*}[t!]
\centering
%\includegraphics[width=60mm,height=40mm]{plots/twidth.pdf}
%\vspace{30mm}
{\includegraphics[width=70mm]{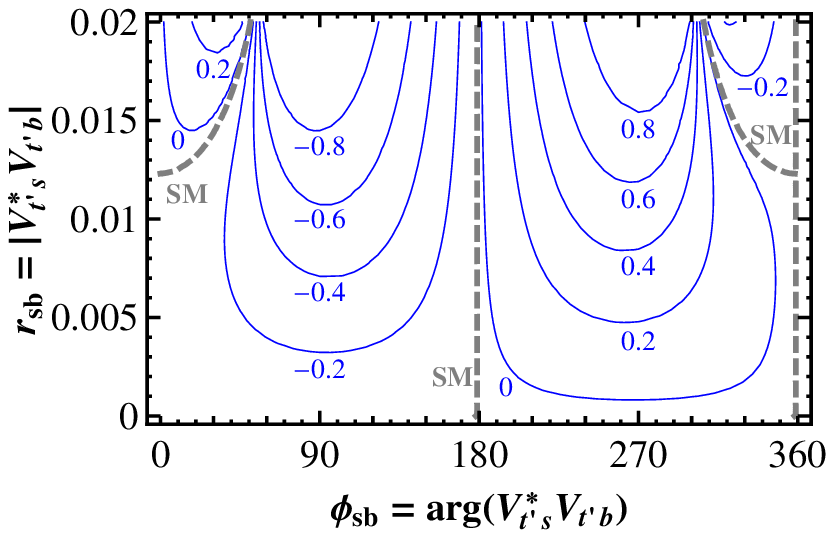}
 \includegraphics[width=70mm]{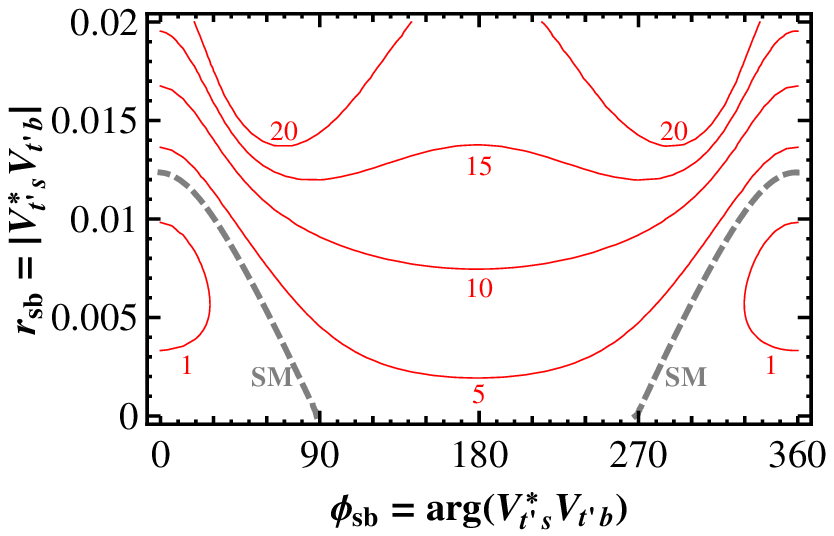}
}
%
%\vskip0.35cm
%{\includegraphics[height=47mm]{plots/tbw.eps}}
%{\includegraphics[width=60mm]{tbw.eps}}
%
\caption{
Contours of (a) $\sin2\Phi_{B_s}$,
 (b) $10^{9}\times {\cal B}(B_s \to \mu^+\mu^-)$
 in $\arg V_{t's}^*V_{t'b}$--$|V_{t's}^*V_{t'b}|$ plane for $m_{t'} = 550$ GeV.
} \label{contours}
\end{figure*}

While $B_s \to \mu^+\mu^-$ has been considered in recent SM4
studies~\cite{Buras:2010pi, Soni:2010xh, Eberhardt:2010bm, Golowich:2011cx}, what we
point out is that, together with the measurements of $\sin2\Phi_{B_s}$ and $m_{t'}$,
the CKM element product $V_{t's}^*V_{t'b}$ can be determined.
Since $V_{us}^*V_{ub}$ and $V_{cs}^*V_{cb}$ are known from tree processes,
a measurement of $V_{t's}^*V_{t'b}$ would already complete the
$b\to s$ quadrangle of Eq.~(1), assuming that one has only SM4
and no other new physics.
This quadrangle %(effectively a triangle as $V_{us}^*V_{ub}$ is rather small)
could be relevant for~\cite{Hou:2008xd} the
baryon asymmetry of our Universe (BAU).
We will discuss the issue of the Higgs boson at the end.

\section{\label{sec:II}\boldmath %Two Measurables and Some
Impact of $\sin2\Phi_{B_s}$ and $B_s \to \mu^+\mu^-$ \protect\\}

The $\bar B_s$--$B_s$ mixing amplitude is well-known,
\begin{eqnarray}
 M_{12}^s &=& \frac{G_F^2M_W^2}{12\pi^2}m_{B_s}f_{B_s}^2\hat B_{B_s}\eta_B
 \Bigl[\left(\lambda_t^{\rm\scriptsize SM}\right)^2 S_0(t,t) \nonumber\\
&& \ \ \ \ \ \ \ \;\;
  + 2\lambda_t^{\rm\scriptsize SM}\lambda_{t'}\Delta S_0^{(1)}
 + \lambda_{t'}^2\Delta S_0^{(2)}
 \Bigr],
 \label{M12s}
\end{eqnarray}
where $\lambda_q \equiv V_{qs}^*V_{qb}$ hence
$-\lambda_t^{\rm\scriptsize SM} = \lambda_c + \lambda_u$, and we have
approximated by factoring out a common
short distance QCD factor $\eta_B$.
With $S_0$ and $\Delta S_0^{(i)}$ as defined in Ref.~\cite{Hou:2006mx},
Eq.~(\ref{M12s}) manifestly respects the Glashow-Iliopoulos-Maiani (GIM) mechanism~\cite{Glashow:1970gm}.

The mass difference $\Delta m_{B_s} \equiv 2|M_{12}^s|$ depends on
the hadronic parameter $f_{B_s}^2\hat B_{B_s}$,
hence it is not useful for extracting short distance information.
However, defining $\Delta_{12}^s = [\; \ldots\; ]$ in Eq.~(\ref{M12s}),
the CPV phase
\begin{equation}
 2\Phi_{B_s} \equiv \arg M_{12}^s = \arg \Delta_{12}^s,
 \label{argM12s}
\end{equation}
depends only on $m_{t'}$ and $\lambda_{t'} = V_{t's}^*V_{t'b}$.
Note that $\lambda_t^{\rm\scriptsize SM} \cong -0.04 -V_{us}^*V_{ub}$,
and we will take the current best fit value for $V_{us}^*V_{ub}$ from PDG~\cite{PDG}.
Note that $V_{us}^*V_{ub}$ can be directly measured via tree processes at LHCb.

We plot, in Fig.~1(a), the contours for $\sin2\Phi_{B_s}$
in the $\phi_{sb} \equiv \arg V_{t's}^*V_{t'b}$,
$r_{sb} \equiv |V_{t's}^*V_{t'b}|$ plane for $m_{t'} = 550$ GeV.
This $m_{t'}$ value is chosen because 500 GeV is almost ruled out,
while going beyond 550 GeV, one is no longer sure of
the numerical accuracy of Eq.~(\ref{M12s}).
That is, above the UB, the perturbative computation
of the functions $\Delta S_0^{(i)}$ would no longer
be valid. However, some form like Eq.~(\ref{M12s})
should continue to hold even above the UB.
We have checked that our results do not change qualitatively
if we straightforwardly apply $m_{t'} = 650$ GeV.

At first sight, the $B_s \to \mu^+\mu^-$ decay rate is also
proportional to $f_{B_s}^2$, bringing in large hadronic
uncertainties. However, this can largely be mitigated~\cite{Buras:2003td}
by taking the ratio with $\Delta m_{B_s}/\Delta m_{B_s}|^{\rm exp}$,
namely
\begin{equation}
\mathcal{B}(B_s\to \bar\mu\mu)
 = C\frac{\tau_{B_s}\eta_Y^2}{\hat{B}_{B_s}\eta_B}
   \frac{|\lambda_t^{\rm\scriptsize SM}Y_0(x_t)+\lambda_{t'}\Delta Y_0|^2}
        {|\Delta_{12}^s|/\Delta  m_{B_s}|^{\rm exp}},
 \label{BrBsmumu}
\end{equation}
where $C = 3g_W^4m_{\mu}^2/2^7\pi^3M_{W}^2$, and
$\eta_Y= \eta_Y(x_t) = \eta_Y(x_{t'})$ is taken.
%
%
%\vspace{1cm}
%The auxiliary functions are
%\begin{align}
%&Y(x)=Y_0(x)+\frac{\alpha_s}{4\pi}Y_1(x)=\eta_Y(x) Y_0(x),\quad  %\eta_Y(x)=\frac{Y(x)}{Y_0(x)}  \\
%&Y_0(x)=\frac{x}{8}\left[\frac{4-x}{1-x}+\frac{3x}{(1-x)^2}\ln %x\right],\nonumber\\
%&Y_1(x)=\frac{10x+10x^2+4x^3}{3(1-x)^2}-\frac{2x-8x^2-x^3-x^4}{(1-x)^3}\ln x %+\frac{2x-14x^2+x^3-x^4}{2(1-x)^3}\ln^2 x\nonumber\\
%&\qquad\qquad +\frac{2x+x^3}{(1-x)^2}\mathrm{Li}_2(1-x)+8x\frac{\partial %Y_0(x)}{\partial x}\ln x_\mu\nonumber\\
%&\mathrm{Li}_2(1-x)=\int^x_1 dt\frac{\ln t}{1-t},\qquad %x_\mu=\frac{\mu^2}{m_{\sw}^2}
%\end{align}
%where $\mu$ is the scale at which the running internal quark mass $m(\mu)$ is %defined.
%
Hadronic dependence is now only in the
better-known ``bag parameter," $\hat B_{B_s}$.
Furthermore, stronger $t'$ dependence is brought in through
the short distance function $|\Delta_{12}^s|$ that enters $\Delta m_{B_s}$.
We plot the contours for ${\cal B}(B_s \to \mu^+\mu^-)$
in the $\phi_{sb}$--$r_{sb}$ plane for $m_{t'} = 550$ in Fig.~1(b).

\begin{figure*}[t!]
\centering
%\includegraphics[width=60mm,height=40mm]{plots/twidth.pdf}
%\vspace{30mm}
{\includegraphics[width=70mm]{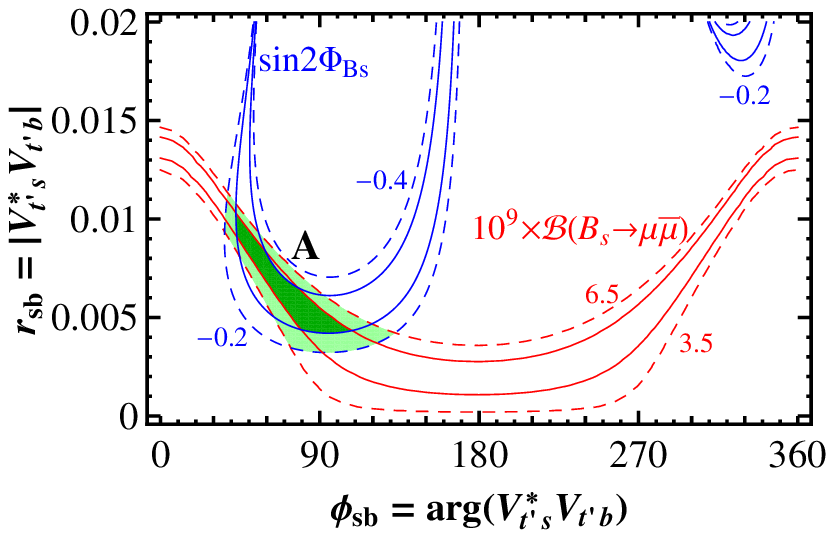}
 \includegraphics[width=70mm]{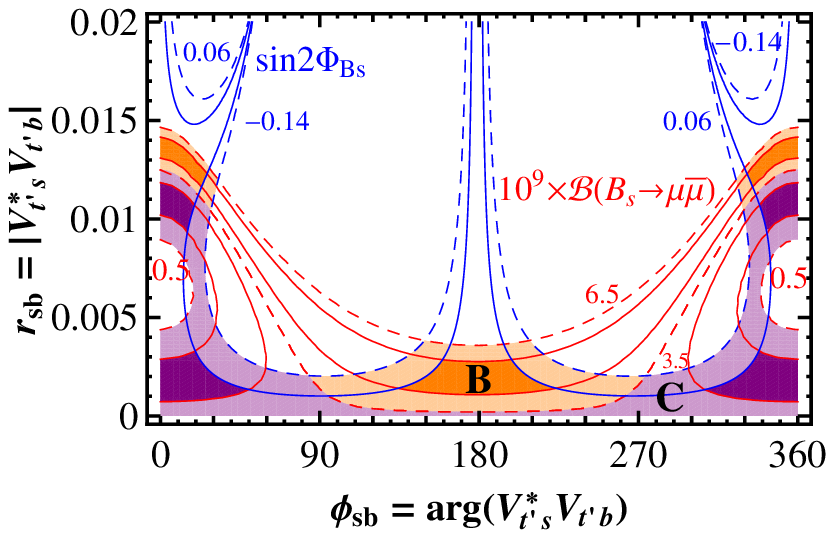}
}
\vskip-0.35cm
\caption{
 Overlap region for (a) $\sin2\Phi_{B_s} = -0.3 \pm 0.1$ and
  $10^{9} \times {\cal B}(B_s \to \mu^+\mu^-) = 5.0 \pm 1.5$ (Case A),
  where solid line is for half the error;
 (b) $\sin2\Phi_{B_s} = -0.04 \pm 0.1$, but
  $10^{9} \times {\cal B}(B_s \to \mu^+\mu^-) = 5.0 \pm 1.5$ (Case B)
  or $2.0 \pm 1.5$ (Case C).
} \label{overlap}
\end{figure*}

To anticipate the progress with full 2011 data, and towards 2012,
we project possible values for $\sin2\Phi_{B_s}$ and
${\cal B}(B_s \to \mu^+\mu^-)$.
The LHCb result of Eq.~(\ref{phis1108}) is at some
odds with earlier results.
A study~\cite{Hou:2010mm} of high mass $m_{t'} = 500$ GeV case
considering all relevant data, as compared with $m_{t'} = 300$ GeV
(now ruled out) case studied earlier~\cite{Hou:2005yb},
suggested a smaller $\sin2\Phi_{B_s}$ value of order $-0.3$.
This value is still within 2$\sigma$ of Eq.~(\ref{phis1108}).
Given the surprise shift from a hint of
a large and negative central value prior to 2011,
the next update could possibly shift back.
Thus, we shall take two possible values
\begin{equation}
\sin2\Phi_{B_s} = -0.3 \pm 0.1;\ -0.04 \pm 0.1 \ \ ({\rm LHCb} > 1\; {\rm fb}^{-1})
\label{sinphis}
\end{equation}
where the first is more aggressive but reflects the past trend,
while the second follows Eq.~(\ref{phis1108}).

An enhanced $\sin2\Phi_{B_s}$ implies the same
for $B_s \to \mu^+\mu^-$, so we should entertain
the possibility that ${\cal B}(B_s \to \mu^+\mu^-)$ is
larger than the SM3 value of $3.2 \times 10^{-9}$.
On the other hand, given that $\sin2\Phi_{B_s}$ is
now suitably consistent with SM3, one should
consider not only the possibility that ${\cal B}(B_s \to \mu^+\mu^-)$
is consistent with SM3, but entertain even the possibility
that ${\cal B}(B_s \to \mu^+\mu^-)$ might be found to be
\emph{less} than the SM3 expectation.
Following the reasoning of Ref.~\cite{Akeroyd:2011kd}
for how the luminosity, hence errors, might scale for
the combination of LHCb and CMS results, we adopt the
two values of
\begin{equation}
10^{9} \, {\cal B}(B_s \to \mu^+\mu^-) = 5.0 \pm 1.5;\ 2.0 \pm 1.5
 \ \ \, ({\rm 2012})
\label{Bsmumu}
\end{equation}
to project into 2012. We have chosen two adjacent regions
of somewhat enhanced vs somewhat suppressed $B_s \to \mu^+\mu^-$,
which contains the SM3 case in intersection.
In the following, we will illustrate with the errors as in Eqs.~(\ref{sinphis})
and (\ref{Bsmumu}), as well as half the error, anticipating further progress
with data.

We illustrate in Fig.~2(a) for $m_{t'} = 550$ GeV the overlap of
the contours for $\sin2\Phi_{B_s}$ and ${\cal B}(B_s \to \mu^+\mu^-)$
%in the $\phi_{sb}$--$r_{sb}$ plane
when both take larger than SM3 values in
Eqs.~(\ref{sinphis}) and (\ref{Bsmumu}).
We denote this as Case A.
The light shaded overlap region correspond to
the 1$\sigma$ range in Eqs.~(\ref{sinphis}) and (\ref{Bsmumu}).
Reducing errors by half, %($\sim$ full 2011-2012 data),
one gets the dark shaded area
by the overlap of the two sets of solid contours.
%Such reduction in error could come,
%for example, by improving with more modes
%for $\sin2\Phi_{B_s}$, or if ATLAS and CMS gain considerably
%more in statistics over LHCb for $B_s \to \mu^+\mu^-$ in 2012.
%
Roughly speaking, the overlap region extends from
$(r_{sb},\; \phi_{sb}) \sim (0.011, 40^\circ)$ to $(0.004, 130^\circ)$.

Figure~2(b) shows the cases when $\sin2\Phi_{B_s} = -0.04 \pm 0.10$
in Eq.~(\ref{sinphis}),
but ${\cal B}(B_s \to \mu^+\mu^-)$ is either higher (Case B)
or lower (Case C) than SM3 expectations in Eq.~(\ref{Bsmumu}).
The shadings are the same as Fig.~2(a).
The two values in Eq.~(\ref{Bsmumu}) complement each other,
as can be seen from Fig.~2(b). Taken together,
Cases B+C complement Case A of Fig.~2(a), where both $\sin2\Phi_{B_s}$
and ${\cal B}(B_s \to \mu^+\mu^-)$ are on the high side.
A remaining Case D is the small region chipped off from Fig.~2(a)
that lies between Case A and Cases B+C.
We do not discuss this case further, as
it can be inferred from Cases A--C.

Inspecting the overlap regions for Cases B and C, both allow
large $r_{sb}$ solutions for $|\phi_{sb}| \lesssim 40^\circ$,
with $r_{sb}$ ranging around 0.013 (0.011) for Case B (C).
There is, however, a low $r_{sb} \lesssim 0.004$ overlap region
for all $\phi_{sb}$, with Cases B and C complementing each other,
with Case B ranging between $90^\circ$ to $270^\circ$.
When $r_{sb}$ is small, in general $\sin2\Phi_{B_s}$
would become close to the SM3 value and become small.
The full domain of $\phi_{sb}$ is allowed, which in turn
has different implications for ${\cal B}(B_s \to \mu^+\mu^-)$.
Note that the contour line of
${\cal B}(B_s \to \mu^+\mu^-) = 3.5 \times 10^{-9}$ is
very close to the SM3 contour of $3.2 \times 10^{-9}$
(the dashed curves in Fig.~1(b)). Thus,
to the left of $90^\circ$ (and to the right of $270^\circ$)
for low $r_{sb}$, ${\cal B}(B_s \to \mu^+\mu^-)$ is
suppressed compared to SM3 (compare Fig.~1(b)), which is precisely Case C.
This is a case that still might emerge at the LHC,
even when $\sin2\Phi_{B_s}$ is found consistent with SM3.
The small $\sin2\Phi_{B_s}$ value can of course turn out
to deviate from SM3 when very high precision is reached.

\begin{figure*}[t!]
\centering
%\includegraphics[width=60mm,height=40mm]{plots/twidth.pdf}
%\vspace{30mm}
{%\includegraphics[width=70mm]{Fig4-AFB_zero.eps} %
 \includegraphics[width=70mm]{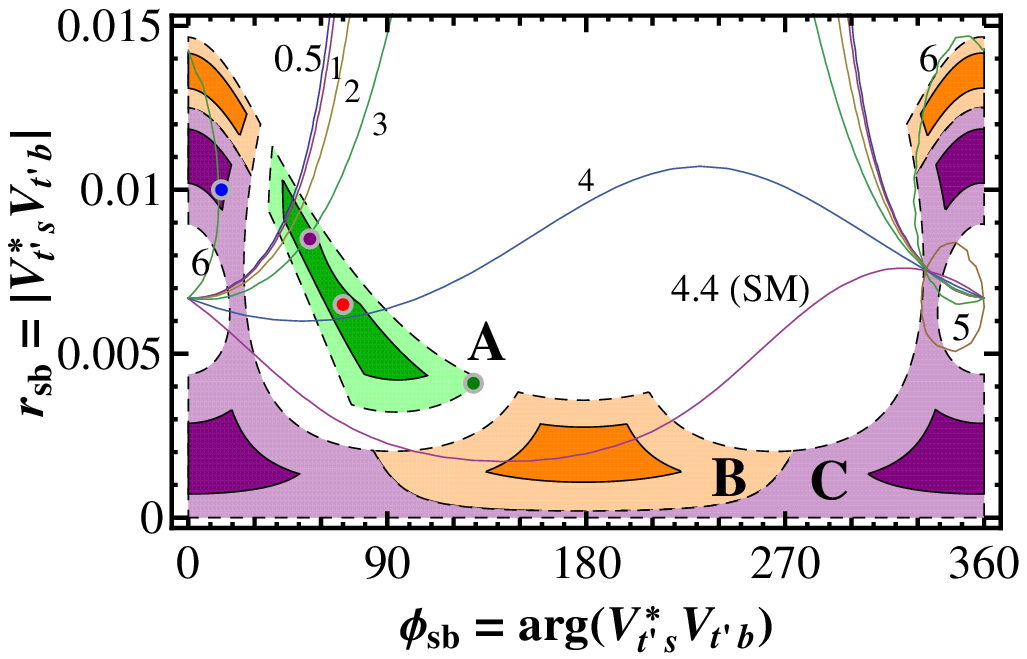}
 \includegraphics[width=70mm]{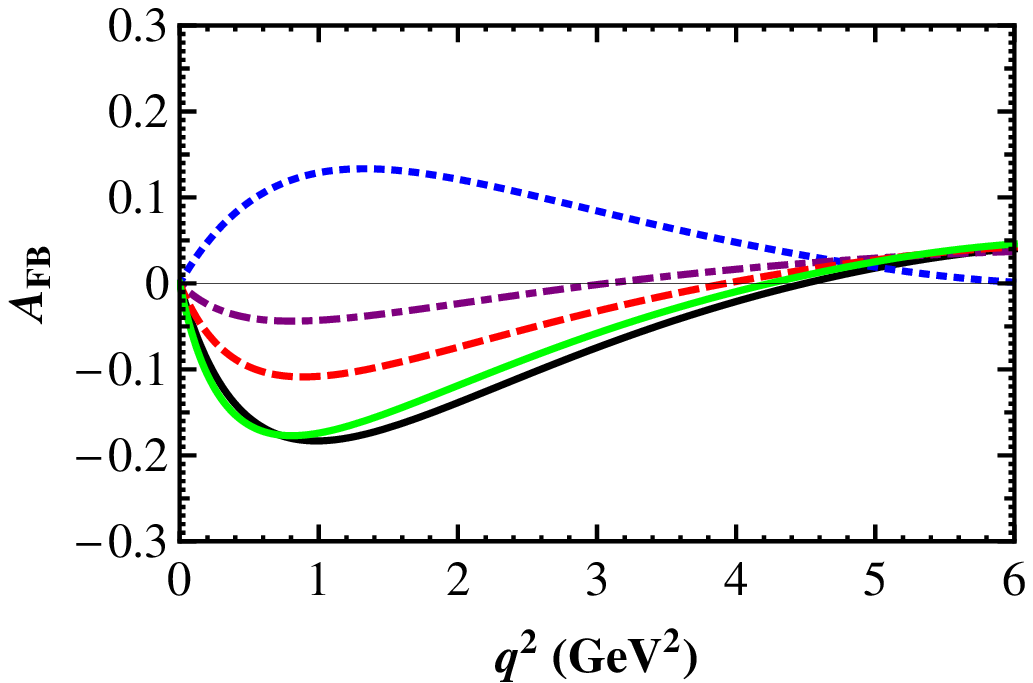}
}
\caption{
(a) Contours of zero crossing $s_0 \equiv q^2|_{A_{\rm FB} = 0}$
    in the $\arg V_{t's}^*V_{t'b}$--$|V_{t's}^*V_{t'b}|$ plane,
    overlayed with the overlap regions of Fig.~2 (for $m_{t'} =$ 550 GeV);
(b) Differential $dA_{\rm FB}/dq^2$ vs $q^2 \equiv m_{\mu^+\mu^-}^2$,
    where curves from top to bottom (the black solid curve is for SM3)
    correspond to sample points in overlap regions in (a),
    indicated by small ellipses, from larger to smaller $r_{sb}$,
    as explained further in text.
} \label{AFB}
\end{figure*}

\section{\label{sec:AFB} \boldmath %A Third Measurable:
Utility of $A_{\rm FB}(B^0 \to K^{*0}\mu^+\mu^-)$\protect\\}

We have focused so far on $\sin2\Phi_{B_s}$ and ${\cal B}(B_s \to \mu^+\mu^-)$,
the two $B$ physics trump cards in the quest for new physics at the LHC.
But a third measurable can be done well by LHCb:
the forward-backward asymmetry in $B^0 \to K^{*0}\mu^+\mu^-$.
Earlier measurements~\cite{PDG} by the B factories, and by CDF,
found no indication of a zero crossing.
%Rather, a sign-flipped (from SM3) $C_7$
%coefficient gives better fit.
%Though the CPV phase of $C_7$ could change mildly~\cite{Hou:2005yb} in SM4,
%it is not enough to flip the sign.
However, the summer 2011 result~\cite{Patel} of LHCb once again
turned out in support of SM3.
This has implications on the overlap regions of Fig. 2.

The zero crossing point $s_0 \equiv q^2|_{A_{\rm FB} = 0}$
is insensitive to form factors, hence
an important probe of possible new physics.
%could differ from SM3 by changes in $C_{9,10}$.
It has been found generally~\cite{Buras:2010pi, Soni:2010xh}
that, once other flavor and CPV data are taken into account,
the variation in $s_0$ for SM4 probably cannot be
distinguished from SM3 within experimental resolution.
 %$\sigma_{s_0} \simeq\ $?? for 1 fb$^{-1}$ data at LHCb.
%
But to investigate the power of LHC data alone,
we plot in Fig.~3(a) the contours of constant $s_0$
in the $\phi_{sb}$--$r_{sb}$ plane for $m_{t'} = 550$,
overlaid with the overlap regions of Fig.~2.
We will now show that the consistency  of the summer 2011
$A_{\rm FB}(B \to K^*\mu^+\mu^-)$ result of LHCb~\cite{Patel} with SM3
rules out the low $\phi_{sb}$, high $r_{sb}$ region,
as well as the upper tip of allowed region for Case A.

We take sample points from the overlap regions,
illustrated as small ellipses in Fig.~3(a),
and plot the corresponding $dA_{\rm FB}/dq^2$ vs
$q^2 \equiv m_{\mu^+\mu^-}^2$ in Fig.~3(b),
where the black solid curve is for SM3.
For the more interesting Case A,
i.e., $\sin2\Phi_{B_s} = -0.3\pm 0.1$ and
${\cal B}(B_s \to \mu^+\mu^-) = (5.0 \pm 1.5) \times 10^{-9}$
both enhanced over SM3 values, we take
\begin{equation}
 V_{t's}^*V_{t'b} \equiv r_{sb}\, e^{i\phi_{sb}} \simeq 0.0065\, e^{i70^\circ},
\label{Vt'sVt'b}
\end{equation}
which lies near the center of the allowed region for Case~A
 (third small ellipse from left in Fig.~3(a)),
and is close to the $s_0 \simeq 4$ GeV$^2$ contour.
This is plotted as the red dashed curve in Fig.~3(b),
where we have used the form factor model of Ref.~\cite{Ball:2004rg}
within QCD factorization framework.
Indeed, the zero crossing lies lower than
the black solid SM3 curve, with $A_{\rm FB}$
weaker than SM3 below the zero crossing.
But away from the zero crossing point,
form factor model dependence would set in,
hence we deem the vicinity of this region in $\phi_{sb}$--$r_{sb}$
as allowed by $A_{\rm FB}$.
If one moves to the lower right tip of Case~A,
one moves closer to $s_0 \simeq 4.4$ GeV$^2$ contour
of Fig.~3(a), hence $A_{\rm FB}$ would be even
harder to distinguish from SM3.
This is illustrated by the green (light grey) solid curve
in Fig.~3(b) for the sample point of
$V_{t's}^*V_{t'b} = 0.004\, e^{i130^\circ}$ (see Fig.~3(a)),
which is indeed hard to distinguish from the SM3 curve.
In fact, it is easily checked that for all points
with $r_{sb} \lesssim 0.004$, $A_{\rm FB}$ would appear
SM3-like.

The opposite is true for large $r_{sb}$ case.
Within Case A, let us take the sample point of
$V_{t's}^*V_{t'b} = 0.0085\, e^{i55^\circ}$, which roughly
sits on the $s_0 \simeq 3$ GeV$^2$ contour of Fig.~3(a)
 (second small ellipse from left),
and is in the upper, darker shaded region for Case A.
This point is plotted as the purple dotdashed line
in Fig.~3(b), with indeed $s_0 \simeq 3$ GeV$^2$.
But now the $A_{\rm FB}$ value is so low for all
$q^2 < 6$ GeV$^2$, LHCb could probably tell it apart,
even with form factor uncertainties.
However, low $A_{\rm FB}$ values would make
the precise determination of $s_0$ harder.
As an extreme case,
we take $V_{t's}^*V_{t'b} = 0.01\, e^{i15^\circ}$
 (first small ellipse from left in Fig.~3(a)),
which is plotted as the blue dotted curve in Fig.~3(b).
This $\phi_{sb}$--$r_{sb}$ combination falls on the
$s_0 \simeq 6$ GeV$^2$ contour in Fig.~3(a),
as we can see also from the $dA_{\rm FB}/dq^2$ plot.
However, $A_{\rm FB}$ now has the \emph{wrong sign} as compared
with data, hence this region is \emph{ruled out}.
This in fact applies to the whole region to the left of, roughly
(to be determined fully by experiment) the $s_0 \simeq 0.5$ GeV$^2$ contour.
Together with the previous point that $s_0 \simeq 3$ GeV$^2$
probably would involve $A_{\rm FB}$ values that are too small,
practically all $r_{sb} \gtrsim 0.008$ regions are
ruled out, or disfavored, by $A_{\rm FB}$ measurement.

A little further explanation can shed light on the $A_{\rm FB}$ behavior.
The differential $dA_{\rm FB}/dq^2$ is proportional to the strength of
the Wilson coefficient $C_{10}$, while the $B_s \to \mu^+\mu^-$ amplitude
is proportional to $C_{10}$. The point of convergence of
the $s_0$ contours in Fig.~3(a) for $\phi_{sb} = 0$ corresponds to
the vanishing point for ${\cal B}(B_s \to \mu^+\mu^-)$.
$C_{10}$ crosses through zero at this point, and has opposite sign
above and below. This explains the sign of the blue dotted curve
in Fig.~3(b).
There is a second convergence point for the $s_0$ contours in Fig.~3(a),
and one could see ellipse shaped contours, e.g. for $s_0 = 5$ GeV$^2$.
This is because $dA_{\rm FB}/dq^2$ is a quadratic function of $r_{sb}\,e^{i\phi_{sb}}$.
One has similar behavior that the upper part of the $s_0 = 5$ GeV$^2$
ellipse give the wrong sign for $A_{\rm FB}$.

%It is clear that a rather low $s_0$, even mimicking sign-flipped $C_7$
%(achieved here with $C_{10}$ turned purely imaginary)
%is possible.
%If this emerges at LHCb (and perhaps ATLAS and CMS too),
%it would suggest that, in the solution space of
%the $\sin2\Phi_{B_s}$ and ${\cal B}(B_s\to \mu^+\mu^-)$ measurements,
%the largest strength for $V_{t's}^*V_{t'b}$,
%together with modest phase angle, may be preferred.
%The experiments can of course study this in more detail
%as data accumulates.

\begin{figure*}[t!]
\centering
%\includegraphics[width=60mm,height=40mm]{plots/twidth.pdf}
%\vspace{30mm}
{\includegraphics[width=115mm]{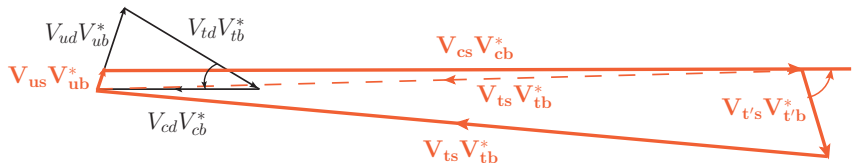}
}
\caption{
The $b\to d$ and $b\to s$ triangles of SM3,
and the $b\to s$ quadrangle of SM4,
with $V_{t's}V_{t'b}^*$ taken from Eq.~(\ref{Vt'sVt'b}).
} \label{bsQUAD}
\end{figure*}

\section{\label{sec:Implication} \boldmath
Implications and Discussion\protect\\}

We would like to give some interpretation of the impact of
this possible future extraction of $\phi_{sb}$ and $r_{sb}$.
We illustrate with the relatively aggressive value of Eq.~(\ref{Vt'sVt'b}),
which corresponds to $\sin2\Phi_{B_s} = -0.3\pm 0.1$ and
${\cal B}(B_s \to \mu^+\mu^-) = (5.0 \pm 1.5) \times 10^{-9}$
both enhanced over SM3 values, and $m_{t'} = 550$ GeV.
We note that Eq.~(\ref{Vt'sVt'b}) is consistent with
the finding of Ref.~\cite{Hou:2010mm}, but if it emerged in 2012,
\emph{the information would be purely from these
two measurements from the LHC},
%the larger values in Eqs.~(\ref{sinphis}) and (\ref{Bsmumu}),
rather than from ``global"
considerations~\cite{Buras:2010pi, Soni:2010xh, Eberhardt:2010bm, Hou:2010mm}.

A measurement like Eq.~(\ref{Vt'sVt'b}) would complete
the unitarity quadrangle of Eq.~(\ref{bsQuad}), assuming,
of course that one only established SM4 but no further new physics.
Let us start by drawing the familiar SM3 $b\to d$ triangle,
$V_{ud}V_{ub}^* + V_{cd}V_{cb}^* + V_{td}V_{tb}^* = 0$,
in Fig.~4.
By standard convention~\cite{PDG},
$-V_{cd}V_{cb}^*$ is real and positive,
$V_{ud}V_{ub}^*$ points above the real axis, while
$V_{td}V_{tb}^*$ points from $V_{ud}V_{ub}^*$ to $-V_{cd}V_{cb}^*$,
giving the familiar apex angle $\beta/\phi_1$,
as indicated.
Switching from $b\to d$ to $b\to s$,
$V_{us}V_{ub}^*$ shrinks by $|V_{us}/V_{ud}| \simeq 0.23$ in length,
but it is in the same direction as $V_{ud}V_{ub}^*$.
The real and positive $V_{cs}V_{cb}^*$ extends
parallel to the real axis from $V_{us}V_{ub}^*$
(most presentations by the experiments misrepresent this), but it
is $|V_{cs}/V_{cd}| \simeq 1/0.22$ times longer than $-V_{cd}V_{cb}^*$.
If $V_{t's}V_{t'b}^* = 0$, then
$V_{ts}V_{tb}^*$ brings one straight back to the origin
(dashed line in Fig.~4),
i.e. $V_{us}V_{ub}^* + V_{cs}V_{cb}^* \equiv -V_{ts}V_{tb}^*|^{\rm SM3}$:
one has a rather squashed SM3 $b\to s$ triangle
with tiny $\Phi_{B_s}|^{\rm SM}$,
but the same area as the $b\to d$ triangle.

But with $V_{t's}V_{t'b}^*$ finite as in Eq.~(\ref{Vt'sVt'b}),
$V_{ts}V_{tb}^*$ would now differ from $V_{ts}V_{tb}^*|^{\rm SM3}$,
and carry a larger CPV phase itself.
The quadrangle of Eq.~(1), as shown in Fig.~4,
would be larger in area than the $b\to d$ or $b\to s$
triangles in SM3 by a factor
$|V_{t's}V_{t'b}^*|/|V_{us}V_{ub}^*| \sim 0.0065/0.00088
\simeq 7$, as the strength of phase angle is similar.

Equation~(\ref{Vt'sVt'b}) corresponds to $\sin2\Phi_{B_s}$ that
is $\sim 2\sigma$ away from the current LHCb central value of
Eq.~(\ref{phis1108}), and may not be realized.
Equation~(\ref{phis1108}) prefers a small $\sin2\Phi_{B_s}$ value.
With the large $r_{sb}$ possibilities
ruled out by $A_{\rm FB}$ as discussed,
one is left with $r_{sb} \equiv |V_{t's}V_{t'b}^*| \lesssim 0.004$,
with $\phi_{sb}$ practically unconstrained at present.
One can picture this in Fig.~4 by reducing the
length of $|V_{t's}V_{t'b}^*|$ by 60\%, and with the
full 360$^\circ$ $\phi_{sb}$ area allowed.
This would probably need more data than 2011-2012 to measure.

The LHCb result~\cite{Raven} for $B_s\to J/\psi\phi$ alone
gave a positive central value of $\sin\phi_s = 0.13$.
If this situation is borne out, we note from Fig.~2(b)
that the branch for small and positive $\phi_{sb}$ is
ruled out by $A_{\rm FB}$. But, depending on what
${\cal B}(B_s\to \mu^+\mu^-)$ value turns up, there is
a strip of allowed domain for $\phi_{sb} \in (200^\circ,\ 330^\circ)$.
Following roughly the $\sin2\Phi_{B_s} = +0.06$ dashed line on
the righthand side of Fig.~2(b), the region above Case B and C
(see also Fig.~3(a)) would be inferred.
Larger $r_{sb}$ values for
$\phi_{sb} \simeq 320^\circ$--$330^\circ$
would again be ruled out by $A_{\rm FB}$, but otherwise
$A_{\rm FB}$ for this region would be quite consistent
with SM3. The $b\to s$ quadrangle could again be easily
drawn, with $r_{sb}$ typically in 0.004 to 0.005 range.

We note here a curiosity. In Fig.~1(a), the dashed curves
correspond to SM3 contours, in the presence of $t'$.
Comparing with Fig.~3, the upper left and right curves
are ruled out by $A_{\rm FB}$. The two vertical dashed
lines in Fig.~1(a) corresponds to $V_{t's}V_{t'b}^*$
being ``parallel" to $V_{ts}V_{tb}^*|^{\rm SM3}$.
The quadrangle of Eq.~(\ref{bsQuad}) would then become
degenerate with SM3 hence have the same area.

%\section{\label{sec:Discussion} Discussion\protect\\}

We now offer a few points for further discussion.

The importance of measuring the
SM4 $b\to s$ quadrangle cannot be overemphasized.
It not only reflects possible new physics discoveries
in $\sin2\Phi_{B_s}$ and $B_s\to \mu^+\mu^-$,
but interpreting via Fig.~4 may relate~\cite{Hou:2008xd}
the measurement to BAU.
%Taking $m_{t'} = 500$ GeV, the analogue of the SM3
%Jarlskog invariant~\cite{Jarlskog85} gains a factor of $10^{14}$.
%
Following the steps of Ref.~\cite{Huet:1994jb},
assuming a first-order phase transition,
the generated BAU seems to be in the right ballpark~\cite{HKK11}.
Of course, Ref.~\cite{Huet:1994jb} may not apply to
heavy $m_{t'}$, but the nontrivial step of
extending the computation into strong Yukawa coupling
may address the other questionable assumption of
order of phase transition.
The problem is too important to be brushed aside
just because of current inadequacies.
We have also checked~\cite{HHX11} that the neutron electric dipole moment
could get enhanced to $10^{-31}$ $e\,$cm order,
but it seems safely below the $10^{-28}$ $e\,$cm
reach of the new generation of experiments,
even with hadronic enhancement.
As for the same-sign dilepton asymmetry uncovered by D0,
although SM4 can give large and negative $\sin2\Phi_{B_s}$,
it cannot affect $b\to c\bar cs$ decay,
and here we await the cross-check by LHCb.

A recent ``global fit" (in contrast to
others~\cite{Buras:2010pi, Soni:2010xh, Eberhardt:2010bm, Hou:2010mm})
of SM4 parameters found a rather
small $|V_{t's}^*V_{t'b}| < 10^{-3}$~\cite{Alok:2010zj}.
This could be due to two inputs:
allowing the central value of 1.04 (which violates unitarity)
for $|V_{cs}|$, with an error of 0.06, may have inadvertently
overconstrained $|V_{t's}|$;
holding to the 2\% lattice error for
$\xi \equiv f_{B_s}^2 \hat B_{B_s} / f_{B_d}^2 \hat B_{B_d}$
(with $\Delta m_{B_s}/\Delta m_{B_d}$ precisely measured)
in their fit,
%At such minuscule values, one should not expect measurements
%to appear anywhere close to the values in Eqs.~(5) and (6).
%
but not allowing the larger values of
Eqs.~(\ref{sinphis}) and (\ref{Bsmumu}) as possible \emph{future} input,
may be too strong a bias. 
We should add that the authors of Ref.~\cite{Alok:2010zj}
did not include the hints for sizable $\sin2\Phi_{B_s}$
into their fits.
%
%But this may be risky.
%As a case in point, in the 2010 update~\cite{Davies:2010ip}
%of the ``rooted staggered fermions" by the HPQCD collaboration
%for the analogous $f_{D_s}$ decay constant, the central
%value for $248.0 \pm 2.5$ MeV shifted by 2$\sigma$ from the
%2008 result~\cite{Follana:2007uv} of $241 \pm 3$ GeV,
%with the astounding error of 1\%.
%
In any event,
looking at Table~III of Ref.~\cite{Alok:2010zj},
it seems unreasonable that $|V_{t's}^*V_{t'b}| < 10^{-3}$,
while $|V_{t'd}^*V_{t'b}| > 10^{-3}$ is allowed,
especially when we are just entering the era
for major progress in $b\to s$ measurements.
A small $|V_{t's}^*V_{t'b}|$ is certainly possible,
but the three measurements stressed in this work
would soon dominate the determination.

Why do we retain the SM3 $b\to d$ triangle,
even when we extend to the SM4 $b\to s$ quadrangle?
This point was addressed in the semiglobal analysis of
Ref.~\cite{Hou:2005yb}. When considering kaon constraints
on $V_{t'd}^*V_{t's}$, a CKM unitarity approach showed
that $V_{t'd}V_{t'b}^*$ and $V_{td}V_{tb}^*$ are
relatively colinear with  $V_{td}V_{tb}^*|^{\rm SM3}$,
and cannot be easily distinguished by the
$\sin2\phi_1/\beta$ measurement. This, in fact,
predated the subsequent realization of some tension in
$B_d$ mixing and/or $\epsilon_K$~\cite{Lunghi:2008aa},
and would require Super B factory and kaon studies
to disentangle.

We have used $m_{t'} = 550$ GeV, which is at the unitarity bound,
for our discussion. This value can be uncovered by direct search by 2012.
If, however, the $t'$ and $b'$ quarks are above the UB,
i.e $m_{t'} \gtrsim 550$ GeV, then the 14 TeV run would be necessary.
However, with the Yukawa coupling turned nonperturbative,
the phenomenology may change~\cite{Enkhbat:2011vp}.
On the other hand, we would definitely learn in the
next two years whether $\sin2\Phi_{B_s}$ and $B_s\to \mu^+\mu^-$
are beyond SM3 expectations.

Finally, we should mentioned that if a Higgs boson with SM3-like
cross section and properties emerge at the LHC, indications
of which could appear by end of 2011, SM4 alone would be
in great difficulty~\cite{Djouadi}.
One would have to extend beyond simple SM4, even if
SM-like $t'$ and $b'$ quarks are found.
On the other hand, the standard Higgs of SM3 itself,
with mass below 600 GeV or so, might
get ruled out by 2012. If such is the case,
then we might enter the heavy--Higgs, heavy--quark world
of SM4~\cite{Enkhbat:2011vp}.
We are in exciting times indeed.

\section{\label{sec:Conclusion} Conclusion\protect\\}

In conclusion, although once again SM3 seems to hold sway,
whether time-dependent CPV in $B_s \to J/\psi \phi$ is considerably
stronger than SM3 expectations will be conclusively settled
with the full 2011--2012 data at LHCb,
while one could discover that $B_s \to \mu^+\mu^-$
is mildly enhanced. If such is the case, we have shown
that the fourth generation $b\to s$ unitarity quadrangle
would become measured, which could have a bearing on
the matter-antimatter asymmetry of the Universe.
The main thrusts in this quest at the LHC are $\sin2\Phi_{B_s}$,
${\cal B}(B_s \to \mu^+\mu^-)$ and $A_{\rm FB}(B^0 \to K^{*0}\mu^+\mu^-)$.

\vskip0.3cm
\noindent{\bf Acknowledgement}.  WSH thanks the National Science
Council for an Academic Summit grant, NSC 100-2745-M-002-002-ASP,
while MK and FX are supported under the NTU grant
10R40044 and the Laurel program.

\vskip0.2cm
\noindent \textbf{Note Added}.
 Immediately after submission of our work,
 we learned that CDF measured~\cite{CDFmumu11}
 ${\cal B}(B_s\to \mu^+\mu^-) = (18^{+11}_{\;-9})\times 10^{-9}$,
 which was countered by lower values from LHCb~\cite{LHCb-mumu11}
 and CMS~\cite{CMSmumu11} \emph{within a week}.
 The subsequent rapid unfolding of the LHCb results of
 $A_{\rm FB}(B^0 \to K^{*0}\mu^+\mu^-)$ at EPS-HEP 2011,
 and $\sin\phi_s$ at LP 2011 was both exhilarating and
 somewhat disappointing, and resulted in
 major revision of this paper.

%We thank NSC and MOE for support.


\begin{thebibliography}{9}   % Use for  1-9  references
%\begin{thebibliography}{99} % Use for 10-99 references

%
\bibitem{PDG}
  K. Nakamura \textit{et al.} [Particle Data Group], J. Phys. G \textbf{37}, 075021 (2010).

%
\bibitem{Kobayashi:1973fv}
  M.~Kobayashi and T.~Maskawa,
%  ``CP Violation In The Renormalizable Theory Of Weak Interaction,''
  Prog.\ Theor.\ Phys.\  {\bf 49}, 652 (1973).

%
\bibitem{Cabibbo:1963yz}
  N.~Cabibbo,
  %``Unitary Symmetry and Leptonic Decays,''
  Phys.\ Rev.\ Lett.\  {\bf 10}, 531 (1963).

%
\bibitem{Glashow:1970gm}
  S.L.~Glashow, J.~Iliopoulos and L.~Maiani,
  %``Weak Interactions with Lepton-Hadron Symmetry,''
  Phys.\ Rev.\  D {\bf 2}, 1285 (1970).

%
\bibitem{BelleNature}
  S.-W.~Lin {\it et al.} %, Y. Unno, W.-S. Hou, P. Chang
  [Belle Collaboration],
  %``Difference in direct charge-parity violation between charged and neutral B
  %meson decays,''
  Nature {\bf 452}, 332 (2008).

%
\bibitem{HNS}
  W.-S.~Hou, M.~Nagashima and A.~Soddu,
%  ``Difference in $B^+$ and $B^0$ direct $CP$ asymmetry as effect of a fourth generation,''
  Phys.\ Rev.\ Lett.\  {\bf 95}, 141601 (2005).
  %[arXiv:hep-ph/0503072];

%
\bibitem{Hou:2005yb}
  W.-S.~Hou, M.~Nagashima and A.~Soddu,
  %``Enhanced K(L) --> pi0 nu anti-nu from direct CP violation in B --> K pi
  %with four generations,''
  Phys.\ Rev.\  D {\bf 72}, 115007 (2005).

%\cite{Hou:2006mx}
\bibitem{Hou:2006mx}
  W.-S.~Hou, M.~Nagashima and A.~Soddu,
  %``Large time-dependent CP violation in B/s0 system and finite D0 - anti-D0
  %mass difference in four generation standard model,''
  Phys.\ Rev.\  D {\bf 76}, 016004 (2007).
  %[arXiv:hep-ph/0610385].

%
\bibitem{LHCb10}
   The LHCb Collaboration, LHCb-CONF-2011-006.

%
\bibitem{Raven}
  Plenary talk by G. Raven at Lepton Photon Symposium, August 2011,
  Mumbai, India.

%
%\bibitem{Teubert}
   %A. Golutvin, presentation at CERN LHC RRB Meeting April 2011,
   %https://indico.cern.ch/conferenceDisplay.py?confId=128046;
%   F. Teubert, experimental summary talk at \emph{Flavor Physics and CP Violation 2011},
%   May 2011, Israel.

%
\bibitem{Kribs:2007nz}
  G.D.~Kribs {\it et al.}, % T.~Plehn, M.~Spannowsky and T.M.P.~Tait,
%  ``Four generations and Higgs physics,''
  Phys.\ Rev.\  D {\bf 76}, 075016 (2007);
  %[arXiv:0706.3718 [hep-ph]].
  H.-J.~He, N.~Polonsky and S.-f.~Su,
  %``Extra families, Higgs spectrum and oblique corrections,''
  Phys.\ Rev.\  D {\bf 64}, 053004 (2001);
  V.A.~Novikov, L.B.~Okun, A.N.~Rozanov and M.I.~Vysotsky,
  %``Mass of the Higgs versus fourth generation masses,''
  JETP Lett.\  {\bf 76}, 127 (2002)
  [Pisma Zh.\ Eksp.\ Teor.\ Fiz.\  {\bf 76}, 158 (2002)].

%
\bibitem{Holdom:2009rf}
  For a recent brief review on the fourth generation, see
  B.~Holdom {\it et al.} %, W.-S.~Hou, T.~Hurth, M.L.~Mangano, S.~Sultansoy and G.~\"{U}nel,
%  ``Four Statements about the Fourth Generation,''
  PMC Phys.\  A {\bf 3}, 4 (2009).
  %[arXiv:0904.4698 [hep-ph]].

%
\bibitem{Chatrchyan:2011em}
  S.~Chatrchyan {\it et al.}  [CMS Collaboration],
  %``Search for a Heavy Bottom-like Quark in pp Collisions at sqrt(s) = 7 TeV,''
  Phys.\ Lett.\  B {\bf 701}, 204 (2011).
  %[arXiv:1102.4746 [hep-ex]].

%
\bibitem{DeRoeck}
  Plenary talk by A. De Roeck at Lepton Photon Symposium, August 2011,
  Mumbai, India.

%
\bibitem{Chanowitz:1978uj}
  M.S.~Chanowitz, M.A.~Furman and I.~Hinchliffe,
  %``Weak Interactions Of Ultraheavy Fermions,''
  Phys.\ Lett.\  B {\bf 78}, 285 (1978).

%
\bibitem{Aaij:2011rj}
  R.~Aaij {\it et al.}  [LHCb Collaboration],
  %``Search for the rare decays Bs -->mumu and Bd -->mumu,''
  Phys.\ Lett.\  B {\bf 699}, 330 (2011).

%
\bibitem{Buras:2010pi}
  A.J.~Buras {\it et al.}, %, B.~Duling, T.~Feldmann, T.~Heidsieck, C.~Promberger and S.~Recksiegel,
  %``Patterns of Flavour Violation in the Presence of a Fourth Generation of
  %Quarks and Leptons,''
  JHEP {\bf 1009}, 106 (2010).

%
\bibitem{comboBsmumu}
  The combined summer 2011 limit of LHCb and CMS on $B_s\to \mu^+\mu^-$
  can be found in the documents LHCb-CONF-2011-043 and CMS PAS BPH-11-019.

%
\bibitem{Soni:2010xh}
  A.~Soni {\it et al.}, %, A.K.~Alok, A.~Giri, R.~Mohanta and S.~Nandi,
  %``SM with four generations: Selected implications for rare B and K decays,''
  Phys.\ Rev.\  D {\bf 82}, 033009 (2010).

%
\bibitem{Eberhardt:2010bm}
  O.~Eberhardt, A.~Lenz and J.~Rohrwild,
  %``Less space for a new family of fermions,''
  Phys.\ Rev.\  D {\bf 82}, 095006 (2010).

%
\bibitem{Golowich:2011cx}
  E.~Golowich {\it et al.}, % J.~Hewett, S.~Pakvasa, A.A.~Petrov and G.K.~Yeghiyan,
  %``Relating $_s Mixing and B_s \to mu+mu- with New Physics,''
  Phys.\ Rev.\  D {\bf 83}, 114017 (2011).

%
\bibitem{Hou:2008xd}
  W.-S.\ Hou,
%  ``CP Violation and Baryogenesis from New Heavy Quarks,''
  Chin.\ J.\ Phys.\ {\bf 47}, 134\ (2009).

%
\bibitem{Buras:2003td}
  A.J.~Buras,
  %``Relations between Delta M(s,d) and B(s,d) ---> mu anti-mu in models with
  %minimal flavor violation,''
  Phys.\ Lett.\  B {\bf 566}, 115 (2003).

%
\bibitem{Hou:2010mm}
  W.-S.~Hou and C.-Y.~Ma,
  %``Flavor and CP Violation with Fourth Generations Revisited,''
  Phys.\ Rev.\  D {\bf 82}, 036002 (2010).

%
\bibitem{Akeroyd:2011kd}
  A.G.~Akeroyd, F.~Mahmoudi and D.M.~Santos,
  %``The decay Bs -> mu+ mu-: updated SUSY constraints and prospects,''
  arXiv:1108.3018. %[hep-ph].

%
%\bibitem{Jarlskog85}
%  C.\ Jarlskog,
%  Phys.\ Rev.\ Lett.\ {\bf 55}, 1039 (1985). %; Z.\ Phys.\ C {\bf 29}, 491 (1985).

%
\bibitem{Patel}
  Talk by M. Patel at EPS-HEP Conference, July 2011, Grenoble, France.

%
\bibitem{Ball:2004rg}
  P.~Ball and R.~Zwicky,
  %``B(D,S) ---> rho, omega, K*, phi decay form-factors from light-cone sum
  %rules revisited,''
  Phys.\ Rev.\  D {\bf 71} (2005) 014029;
  M.~Beneke, T.~Feldmann and D.~Seidel,
  %``Systematic approach to exclusive B ---> V l+ l-, V gamma decays,''
  Nucl.\ Phys.\  B {\bf 612} (2001) 25.

%
\bibitem{Huet:1994jb}
  P.~Huet and E.~Sather,
  %``Electroweak baryogenesis and standard model CP violation,''
  Phys.\ Rev.\  D {\bf 51}, 379 (1995).

\bibitem{HKK11}
  W.-S. Hou, Y. Kikukawa and M. Kohda,
  unpublished.

\bibitem{HHX11}
  J. Hisano, W.-S. Hou and F. Xu,
  arXiv:1107.3642 [Phys.\ Rev.\ D (to be published)].

%
\bibitem{Alok:2010zj}
  A.K.~Alok, A.~Dighe and D.~London,
  %``Constraints on the Four-Generation Quark Mixing Matrix from a Fit to
  %Flavor-Physics Data,''
  Phys.\ Rev.\  D {\bf 83}, 073008 (2011).

%
%\bibitem{Davies:2010ip}
%  C.T.H.~Davies {\it et al.}, % C.~McNeile, E.~Follana, G.P.~Lepage, H.~Na and J.~Shigemitsu,
  %``Update: Precision D_s decay constant from full lattice QCD using very fine
  %lattices,''
%  Phys.\ Rev.\  D {\bf 82}, 114504 (2010).

%
%\bibitem{Follana:2007uv}
%  E.~Follana {\it et al.}, % C.T.H.~Davies, G.P.~Lepage and J.~Shigemitsu
  %[HPQCD and UKQCD Collaborations],
  %``High Precision determination of the pi, K, D and D(s) decay constants from
  %lattice QCD,''
%  Phys.\ Rev.\ Lett.\  {\bf 100}, 062002 (2008).

%
\bibitem{Lunghi:2008aa}
  E.~Lunghi and A.~Soni,
  %``Possible Indications of New Physics in B(d)-mixing and in sin(2 beta)
  %Determinations,''
  Phys.\ Lett.\  B {\bf 666}, 162 (2008);
  A.J.~Buras and D.~Guadagnoli,
  %``Correlations among new CP violating effects in $\Delta$ F = 2
  %observables,''
  Phys.\ Rev.\  D {\bf 78}, 033005 (2008).

%
\bibitem{Enkhbat:2011vp}
  See, for example, the discussion by T.~Enkhbat, W.-S.~Hou and H.~Yokoya,
  %``Early LHC Phenomenology of Yukawa-bound Heavy Q\bar{Q} Mesons,''
  arXiv:1109.3382. %[hep-ph].

%
\bibitem{Djouadi}
  Plenary talk by A. Djouadi at Lepton Photon Symposium, August 2011,
  Mumbai, India.

%
\bibitem{CDFmumu11}
  %T.~Altonen {\it et al.} [CDF Collaboration], arXiv:1107.2304 [hep-ex],
  %to appear in Phys.\ Rev.\ Lett.
  T.~Aaltonen {\it et al.} [CDF Collaboration],
  %``Search for $B_s \to \mu^+\mu^-$ and $B_d \to \mu^+\mu^-$ Decays with CDF II,''
  Phys.\ Rev.\ Lett.\  {\bf 107}, 191801 (2011).
  %[arXiv:1107.2304 [hep-ex]].

%
\bibitem{LHCb-mumu11}
  Talk by J. Serrano at EPS-HEP Conference, July 2011, Grenoble, France.

%
\bibitem{CMSmumu11}
  S.~Chatrchyan {\it et al.}  [CMS Collaboration],
  %``Search for B(s) and B to dimuon decays in pp collisions at 7 TeV,''
  Phys.\ Rev.\ Lett.\  {\bf 107}, 191802 (2011).
  %arXiv:1107.5834 [hep-ex].

\end{thebibliography}
\end{document}